\documentclass[journal=jacsat,manuscript=article]{achemso}

\usepackage{graphicx}
\usepackage{graphics}
\usepackage{epsfig} 
\usepackage{color}   
\usepackage{hyperref} 

\usepackage{bm}

\graphicspath{{Figures/}} 
\usepackage{xcolor}
\usepackage{dcolumn}
\usepackage{gensymb} 

\usepackage{xr}

\title{Coupled metamaterial-phonon terahertz range polaritons in a topological insulator}

\author{Sirak M. Mekonen}
\affiliation{William H. Miller III , Department of Physics and Astronomy, The Johns Hopkins University, Baltimore, MD, 21218, USA} 
\author {Deepti Jain}
\affiliation{Department of Physics and Astronomy, Rutgers, The State University of New Jersey, Piscataway, NJ 08854, USA}
\affiliation{Center for Quantum Materials Synthesis, Piscataway, NJ 08854, USA}
\author{Seongshik Oh}
\affiliation{Department of Physics and Astronomy, Rutgers, The State University of New Jersey, Piscataway, NJ 08854, USA}
\affiliation{Center for Quantum Materials Synthesis, Piscataway, NJ 08854, USA}
\author{N.P. Armitage}
\affiliation{William H. Miller III , Department of Physics and Astronomy, The Johns Hopkins University, Baltimore, MD, 21218, USA} 
\email{npa@jhu.edu; smm@jhu.edu}

\date{\today}

\begin{document}

\begin{abstract}

We report terahertz time-domain spectroscopy (TDTS) experiments demonstrating strong light-matter coupling in a terahertz (THz) LC-metamaterial in which the phonon resonance of a topological insulator (TI) thin film is coupled to the photonic modes of an array of electronic split-ring resonators. As we tune the metamaterial resonance frequency through the frequency of the low frequency $\alpha$ mode of (Bi$_x$Sb$_{1-x}$)$_2$Te$_3$ (BST), we observe strong mixing and level repulsion between phonon and metamaterial resonance. This hybrid resonance is a phonon polariton. We observe a normalized coupling strength, $\eta$ = $\Omega_R$/$\omega_c$ $\approx$ 0.09, using the measured vacuum Rabi frequency and cavity resonance.  Our results demonstrate that one can tune the mechanical properties of these materials by changing their electromagnetic environment and therefore modify their magnetic and topological degrees of freedom via coupling to the lattice in this fashion.
\end{abstract}
\maketitle

\section{Introduction}

Metamaterials (MMs) are artificial composite systems that offer exceptional control of electromagnetic properties due to the capability to engineer their electric and magnetic resonances by controlling the geometry and size of the individual subwavelength constituents. They offer the possibility to achieve strong coupling between highly confined electromagnetic fields and localized or propagating quasiparticles such as surface plasmon polaritons in metals and superconductors~\cite{gramotnev2010plasmonics}, phonon polaritons in polar dielectrics~\cite{shelton2011strong,kim2020phonon}, and exciton polaritons in organic molecules and transition metal dichalcogenides~\cite{as2022mie,dintinger2005strong, ramezani2017plasmon}.  Recently, MMs have been used to control the electron-phonon interaction of topological insulators via their surface states~\cite{in2018control}. Topological insulators (TIs), a class of quantum materials with robust metallic surface states protected by the topological properties of the bulk wavefunctions~\cite{hasan2010colloquium, ando2013topological, wu2013sudden,autore2017terahertz}, have gathered a growing interest due to both their interesting fundamental physics as well as potential applications in terahertz (THz) detectors~\cite{zhang2010topological} and spintronic devices~\cite{chen2009experimental}. These applications can potentially be realized through the polariton interaction, which arises from strong light-matter interactions between confined electromagnetic field (or cavity resonance) and a matter excitation.

A strong light-matter interaction between lattice vibration and a confined electromagnetic field can reach the strong coupling regime where coherent exchange of energy between light and matter becomes reversible. In this regime, coupled light-matter polaritons form  hybrid states where they can coherently exchange energy at the characteristic rate of the vacuum Rabi frequency $\Omega_R$, which is dominant with respect to other loss mechanisms in the system~\cite{dovzhenko2018light, benz2015control}. A polariton system based on novel functional materials could offer an efficient quantum level system  with tunable sources and detectors, optical filters and qubits operating in the far-infrared frequency range~\cite{bakker1994investigation,tanabe2003frequency, kojima2003far,jin2019modifying,ohtani2019electrically}. They may also afford the possibility of tuning the mechanical properties of materials (and therefore their electronic or magnetic properties through phonon coupling) by changing their electromagnetic environment. Phonon-polariton coupled systems with metamaterials in the mid-infrared range~\cite{shelton2011strong, pons2019launching} have been shown in previous investigations as well as THz range surface-plasmon polaritons ~\cite{liu2015highly,maier2006terahertz,liang2015chip}. In the most dramatic cases, it has been proposed that one can drive phase transitions in materials like SrTiO$_3$ via cavity coupling~\cite{latini2021ferroelectric}.
In this work, we present evidence of strong coupling between the $\alpha$ phonon mode in (Bi$_x$Sb$_{1-x}$)$_2$Te$_3$ (BST) thin films and the inductive-capacitive (\textit{LC}) resonance of a split ring resonator (SRR) metasurface via the emergence of level repulsion.  We performed time-domain terahertz spectroscopy (TDTS) on the resulting BST-SRR hybrid metasurfaces.  The observed level repulsion results in the opening of a small transmission window within the absorption band of the uncoupled phonon. We expect the strength of the coupling can be altered through the design of the SRR. Mode assignments were aided through extensive simulations.  In order to parametrically sweep the \textit{LC} resonance frequency across the $\alpha$ mode, we fabricated multiple metasurfaces using standard photolithography techniques. Our measurements reveal level repulsion and hybridization of the coupled systems. This was evident by the formation of a large Rabi splitting with a normalized coupling strength $\eta \approx 0.09$. Our result is the first to show the control over the mechanical properties of TIs by tuning their electromagnetic environment in the THz frequency range. 
 
\begin{figure}[t!]
   \includegraphics[width=0.6\columnwidth]{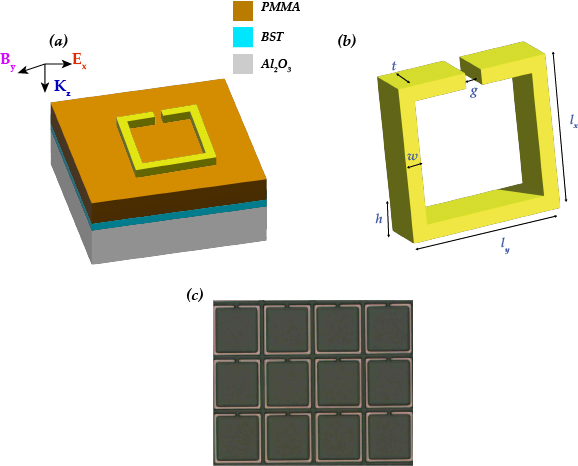} 
   \caption{ {Design of multiscale topological insulator metasurfaces.} a) Schematic of the unit cell showing thin film interface between metallic SRR and Al$_2$O$_3$ substrate with the corresponding electromagnetic excitation configuration.  b) Side-view schematic of the unit cell with the relevant periods $p$ and resonantor size $l$: \textit{$p_x = p_y$} = 44 $\mu$m, \textit{l} = 34 $\mu$m, \textit{w} = 3 $\mu$m \textit{g} = 1.5 $\mu$m, and \textit{t} = 100 nm. c) An optical microscope image (20x) of a fabricated BST-SRR array.}
\label{fig1}
\end{figure}

\section{Materials and Methods}

Fig. \ref{fig1}(a) is a schematic of the unit cell of our BST film metasurface, which is composed of an array of SRRs deposited on a TI film. The gap in SRRs serve as a capacitor whereas the ring serves as an inductor giving an \textit{LC} resonance. Generally, the resonance frequency of SRRs can be given as \textit{f$_0$} $\approx 1/ (2\pi \sqrt{L_cC})$, where the inductance \textit{$L_c$} and the capacitance \textit{C} are determined by the SRRs dimensions and the effective refractive index of the environment. At the \textit{LC} resonance, the incident electric field induces a large accumulation of oscillating surface charges at the ends of the metal strips resulting in a strong electric field confinement in the capacitive gaps~\cite{pendry1999magnetism,zhang2021ultrastrong,kim2020phonon,chen2006active,kim2018colossal}. The resonance frequency of the SRR generally scales inversely with its dimension. 
We synthesized samples of (Bi$_x$Sb$_{1-x}$)$_2$Te$_3$ (BST) thin films with 20 quintuple layers (QLs) on 0.5 mm thick sapphire (Al$_2$O$_3$) substrate by molecular beam epitaxy as discussed in Ref.~\cite{yao2021suppressing}. We performed finite element method (FEM) based simulations (Ansoft HFSS) to identify the dimensions of SRRs with expected resonance frequencies ranging from 1.1-1.9 THz. Afterwards, the desired metasurfaces were achieved by fabricating SRRs using standard photolithography. We fabricated on top of a 1 $\mu$m PMMA layer on top of a Al$_2$O$_3$ substrate that allowed us -- in addition to presumably tuning the coupling to the film -- to strip off the SRR cleanly and redeposit another configuration of SRRs on top.  The period \textit{$p_x = p_y$} and length \textit{$l_x = l_y$} of the SRRs were varied from 50-44 $\mu$m and 40-28 $\mu$m, respectively, while the width \textit{w} = 3 $\mu$m and gap \textit{g} = 1.5 $\mu$m were fixed. We used a TOPTICA (Teraflash) TDTS system to measure the THz transmission spectra of our samples. The incident THz pulse was polarized parallel to the \textit{x} axis as shown in Fig. \ref{fig1}(a) (See Supplementary Material (SM)).
Fig. \ref{fig1}(c) shows an image of one of the fabricated composite BST-SRR arrays that gives a SRR resonance frequency of 1.5 THz. 

\section{Results and Discussion}

\begin{figure}[t!]
    \includegraphics[width=0.6\columnwidth]{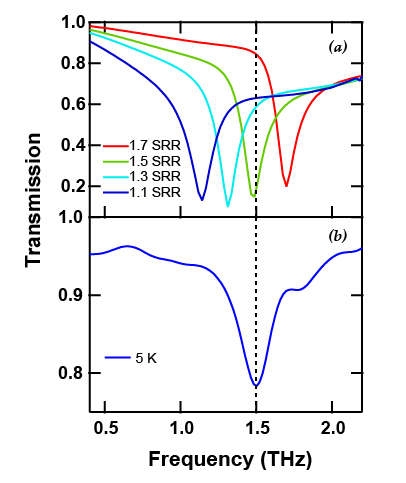}
    \caption{{Numerical simulations of SRRs and TDTS data on Al$_2$O$_3$ at 5K.} a) Simulated THz transmission spectra of SRR's at different resonance frequencies at different \textit{l}. b) TDTS transmission spectrum of the phonon resonance frequency of BST.}
    \label{fig2}
\end{figure}

In Fig. \ref{fig2}(a), we show a FEM simulation of a transmission spectrum for SRRs on a Al$_2$O$_3$ substrate. By tuning the lateral dimension \textit{l} of the SRR, we expect to tune their resonant frequencies from 1.1 (4.55) - 1.7 (7.03) THz (meV). As the dimension of the SRRs decreased, the absorption exhibits a blueshift. Thus, it is possible to match the uncoupled resonant frequency of a SRR to a resonance of material system. The 5K transmission spectra of a bare BST films (with no SRR) is shown in Fig. \ref{fig2}(b). The absorption peak at $\approx$ 1.5 THz is the transverse optical (TO) $\alpha$ phonon mode that in the binary compounds Bi$_2$Se$_3$, Bi$_2$Te$_3$, Sb$_2$Te$_3$ and Sb$_2$Se$_3$, is attributed to an $E_1^u$ mode that corresponds to sliding motion of atomic layers past each other~\cite{richter1977raman}. In the non-stochiometric compounds, phonons in this spectral range were found to extrapolate smoothly with atomic mass from Bi$_2$Te$_3$ to Sb$_2$Te$_3$ and from Bi$_2$Se$_3$ to In$_2$Te$_3$. This $\alpha$ mode has been investigated extensively in the context of THz studies of topological insulators~\cite{wu2013sudden}.

To sweep the SRR frequency across \textit{$\omega_{Ph}$}, we fabricated seven distinct SRRs on 20 nm thick BST films (see SM). At room temperature, the absorption of the SRRs predominates over the phonon mode absorption, primarily because the TI phonon modes become very broad due to scattering, resulting in on overdamped absorption spectra. Consequently, we can accurately predict the frequencies of the SRRs decoupled from phonons, but when deposited on BST, phonons are heavily damped at elevated temperatures (see SM). Therefore, we designate the different SRRs based on these predicted "bare frequencies."

\begin{figure}[t!]
    \includegraphics[width=0.6\columnwidth]{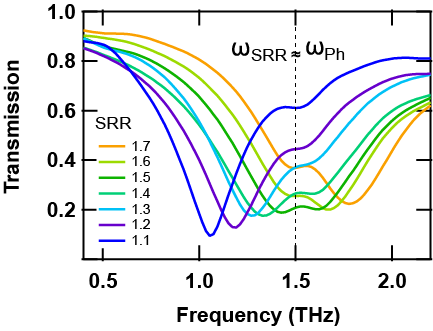}
    \caption{{TDTS data on BST-SRR at 5K} a) TDTS transmission spectrum of the seven SRR metamaterial arrays deposited on the BST films. }
    \label{fig3}
\end{figure}

At low temperatures the behavior is very different. Here the phonon resonance is strong and when the SRR frequency is tuned to it, the effects of mixing and level repulsion are prominent. In Fig. \ref{fig3}, we show the transmission data for the BST-SRR hybrid systems at 5K for \textit{$\omega_{SRR}$} ranging from 1.1 (4.55) to 1.7 (7.03) THz (meV). One can see two notable transmission dips for all samples that indicate two resonances. When their frequencies are far from each other, we can assign a clear local character. Judging from the data in Fig. \ref{fig2}, the more prominent feature has largely SRR character and higher Q-factor. We note though that as the SRR resonance is swept across \textit{$\omega_{Ph}$}, the two peaks always maintain a separation and their intensities become similar. As the resonances are tuned through each other, the lower peak gets further damped and the upper peak sharpens indicating that the local character of excitations change as they are tuned through each other.

We fit the data of Fig. \ref{fig3}, to a double-Lorentizan model to extract the eigenfrequencies $\omega_{-}$ and $\omega_{+}$, and damping rates, $\Gamma_{\pm}$ for all SRRs. Representative fits to these spectra can be found in the supplementary material. Our BST-SRR hybrid system can be considered as two coupled oscillators, one of which has a fixed frequency (the BST phonon) where its electromagnetic environment is tuned by the SRR frequency. When both oscillators are similar frequencies, they form a coupled system and an anticrossing is observed. This results in a periodic transfer of energy between the phonon and SRR through vacuum Rabi oscillations, which is proportional to the splitting at the anticrossing point~\cite{pal2015strong}. 

\begin{figure}[t!]
    \includegraphics[width=0.6\columnwidth]{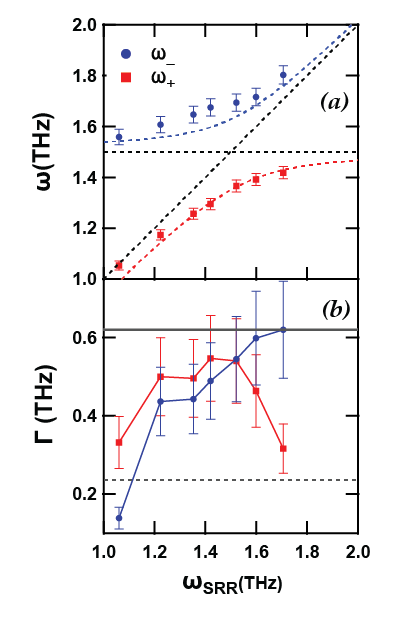}
    \caption{{Dispersion and inverse lifetime of the coupled metamaterial-phonon system.} a) Frequency of hybridized phonon modes of BST and the LC mode of SRRs. b) Inverse lifetime of the resonances.  Grey lines represent the estimation of the uncouple SRR decay rate and the phonon without SRR rates. }
    \label{fig4}
\end{figure}

In Fig. \ref{fig4}(a), we plot the measured eigenfrequencies at 5K versus the uncoupled resonance frequencies that we  obtained at 297K of the BST-SRR hybrid systems. The experimentally obtained peak positions, $\omega_{+}$ and $\omega_{-}$ are shown in circles. One sees a classic signature of level repulsion and mixing of two excitations branches with each other. As the uncoupled resonances approach each other their distinct local characters are lost and new coupled excitations are formed that are symmetric and anti-symmetric combinations of the bare excitation. Our observation is evidence for the formation of a phonon-polariton hybrid from the coupling of the SRR resonance and $\alpha$ mode phonon.

The observed level repulsion behavior can be understood classically and be described using a coupled oscillators model~\cite{novotny2010strong}:
 
\begin{equation}
    \centering 
    \omega_{\pm}^2 = \frac{1}{2}\left[\omega^2_{SRR} + \omega^2_{Ph} \pm \sqrt{(\omega^2_{SRR}-\omega^2_{Ph})^2 +  \Omega_R^2 \omega_{SRR}\omega_{Ph}}\right].
     \label{eq1}
 \end{equation}

We fit the observed level repulsion to Eq.\ref{eq1} as indicated in Fig. \ref{fig4}(a) to obtain the coupling parameters. The strength of the coupling, $\Omega_R$, for when $\omega_{SRR}$ $\approx$ $\omega_{Ph}$, is found to be 0.27 (1.12) THz (meV). The observed splitting is a significant fraction of the $\alpha$ phonon mode resonance which indicates a strong light-matter interaction at the avoided crossing. The normalized coupling strength ratio, $\eta$ = $\frac{\Omega_R}{\omega_{c}}$, between the Rabi frequency and BST-SRR resonance frequency, is found to be $\eta$ $\approx$ 0.09.

It is also interesting to note the behavior of the peak widths as the bare SRR frequency is swept. Fig. \ref{fig4}(b) shows the rates as a function of $\omega_{SRR}$. The solid gray line denotes the SRR-like widths we extrapolated from the data to obtain a reasonable estimate of the width of the uncoupled SRR. As for the phonon width, represented by the dashed gray line, we used the width of the phonon observed in the low-temperature spectra for a film without SRRs. One can see that when $\omega_{SRR}$ is small $\omega_-$ has a lower damping than $\omega_+$ showing its principle SRR character. Near the crossing, the lifetimes are equal showing the mixed character. For large SRR frequency, $\omega_+$, has the smallest damping, shows that now it has largely SRR character (and $\omega_-$ has largely phonon character). It is unclear as to why the $\omega_+$ data for 1.1 THz are below the asymptote of 0.62 THz. We speculate that there is some overall changes to the profile of the electromagnetic fields as the SRR frequency is tuned.

\section{Conclusions}

In this work, we have demonstrated strong light-matter coupling between the $\alpha$ phonon mode of (Bi$_x$Sb$_{1-x}$)$_2$Te$_3$ and cavity resonances of planar THz range metamaterials. We have given spectroscopic evidence of strong coupling with a normalized coupling strength of $\eta$ $\approx$ 0.09. Consequently, we have observed the formation of THz phonon-polariton resonance emerging from the integration of metamaterials with topological insulators.
Our findings hold promise for the advancement of TI-based electronics and plasmonic applications. We anticipate the possibility of transitioning into the regime of ultrastrong coupling for TI. Our work may facilitate the capture of photons for various applications such as slow-light phenomena, quantum computing, and infrared light harvesting. By varying the metamaterial resonance, we have demonstrated the ability to manipulate the mechanical properties of a material by tuning its electromagnetic environment. Via their coupling to phonons this may be used to control magnetic and topological degrees of freedom.

\medskip

\section{Supporting Information}
Our supplementary materials offer further details on the growth of the topological insulator (bismuth antimony telluride), the finite element simulation employed to derive the electromagnetic behavior of the split-ring resonators, the fabrication methods for creating the BST-SRR systems, the spectroscopic technique applied to assess the fabricated devices, and the methods used to interpret the optical response of the systems.

\section{Acknowledgments}
Work at JHU was supported by NSF DMR-1905519 and an AGEP supplement. Work at Rutgers was supported by ARO-W911NF2010108 and MURI W911NF2020166. The work reported here was partially carried out in the Nanofabrication
Facility at the University of Delaware (UDNF). We thank A. Jackson, K. Katsumi, and L.Y. Shi for helpful discussions.


\medskip

\section{Author Contributions}
SMM performed the simulation, fabrication and TDTS measurements. DJ grew the thin films.  SO and NPA supervised the project.  SMM and NPA wrote the manuscript with input from other authors.

\medskip
The authors declare no competing interests.

\medskip
\section{Data Availability Statement}

Source data are available for this paper. All other data that support the plots within this paper and other findings of this study are available from the corresponding author upon reasonable request.

\bibliography{BSTSRR}

\bigskip


\end{document}


\section{\bf{SA}: Growth of Bismuth Antimony Telluride }

The bismuth antimony telluride (Bi$_x$Sb$_{1-x}$)$_2$Te$_3$ (BST), x = 0.14, films were grown on 10 x 10 mm sapphire (Al$_2$O$_3$) (0001) substrates using a custom-built MBE system (SVTA) with base pressure less than 5 × 10$^{-10}$ Torr. Prior to growth, substrates were cleaned ex-situ by UV generated ozone followed by in-situ heating up to 750\degree  under oxygen pressure of 5 × 10$^{-7}$ Torr. All source fluxes (Bi, Sb and Te) were calibrated in-situ with a quartz crystal microbalance (QCM) and ex-situ with Rutherford backscattering spectroscopy (RBS). BST films were grown on a buffer layer of chromium oxide, Cr$_2$O$_3$, and their growth was monitored in-situ using reflection high-energy electron diffraction (RHEED). To protect the topological surface state, the samples were capped after growth with an amorphous 4 nm of Cr$_2$O$_3$. For further details please see Ref. \cite{yao2021suppressing}.

\section{\bf{SB}: Finite Element Method Simulation of Split-Ring Resonators}

\begin{figure}[h!]
    \includegraphics[width=0.4\columnwidth]{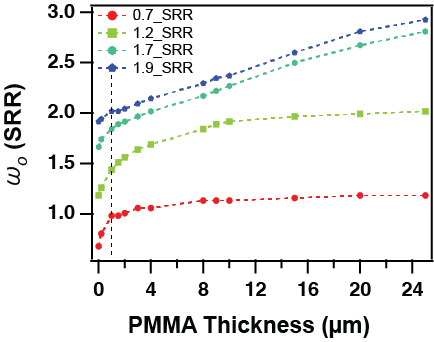}
    \caption{Simulated resonance frequencies for different SRRs for different PMMA thicknesses and different structure configurations.   }
    \label{SFig1}
    
\end{figure}

\begin{figure}[h!]
    \includegraphics[width=0.4\columnwidth]{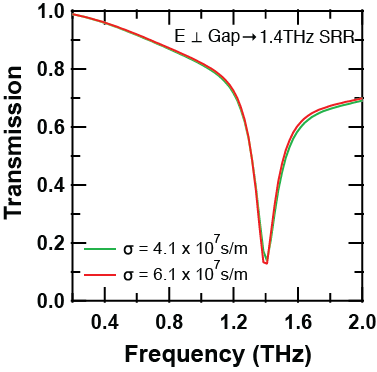}
    \caption{Simulated resonance frequencies of a SRR ($
    \approx$ 1.4 THz) for different conductivity value of gold.}
    \label{SFig5}
    
\end{figure}

Prior to fabrication, we conducted numerical calculations to determine suitable geometries for split-ring resonators (SRRs) with expected resonance frequencies ranging from 1.1 to 1.9 THz. Additionally, as depicted in Fig.\ref{SFig1}, numerical simulations were carried out to explore the impact of PMMA thickness across the same frequency range. As PMMA has smaller index of refraction than sapphire, our expectation was that thicker PMMA layers will have higher resonance frequencies. This was born out by the simulations. Moreover, the results indicate that, for all investigated SRRs, the PMMA layer saturated at $\approx1 \mu$m. Furthermore, we conducted simulations to explore the conductivity dependence, simulating the behavior of gold at low temperatures. It's anticipated that the performance of the SRRs will remain consistent at lower temperatures due to their optical response being minimally affected by the conductivity of the metal they are  composed of, as well as the dissipative environment they inhabit. This assertion is supported by the findings illustrated in Fig.\ref{SFig5}. The simulation calculations were performed using a finite element method (FEM) simulation (Ansoft HFSS). In these simulations, the elementary unit cell of the designed metasurfaces was illuminated at normal incidence, with TM polarization (E $\parallel$ \textit{x} axis). Periodic boundary conditions were applied to mimic a 2D finite structure. The sapphire (Al$_2$O$_3$) substrate was treated as a dielectric with $\epsilon_{sub}$ = 10, the PMMA layer was modeled as a dielectric with $\epsilon_b$ = 1.2, and gold (Au) was modeled as a lossy metal with a conductivity of 4.1 x 10$^7$ S/m.  Initially, these simulations only considered Al$_2$O$_3$ as a dielectric substrate. However, the introduction of the PMMA layer led to a blueshift in the resonance frequency of the SRRs as expected. Consequently, the simulation parameters and dimensions of the SRRs were adjusted to achieve the desired frequency ranges, as outlined in Table \ref{Table 1}.


\begin{table}[h!]
\centering
\begin{tabular}{ |p{2cm}|p{2cm}|p{2cm}|p{2cm}| p{2cm}|}
\hline
SRR (THz) & Period ($\mu$m) & Length ($\mu$m)   & Width ($\mu$m)  & Gap ($\mu$m) \\
\hline
1.1 & 50 &40 &36 &1.5\\
1.2 & 48 &38 &32 &1.5\\
1.3 & 46 &34 &28 &1.5\\
1.4 & 44 &32 &26 &1.5\\
1.5 & 44 &34 &28 &1.5\\
1.6 & 44 &30 &24 &1.5\\
1.7 & 44 &28 &22 &1.5\\
\hline

\end{tabular}

\caption{Simulated parameters for fabricated split ring resonators.}
\label{Table 1}
\end{table}

\section{\bf{SC}: Fabrication of (Bi$_x$Sb$_{1-x}$)$_2$Te$_3$-Split Ring Resonators}

The (Bi$_x$Sb$_{1-x}$)$_2$Te$_3$-Split Ring Resonators (BST-SRR) arrays were fabricated using a standard photolithography method and  etch: a 1 $\mu$m thick polymethyl methacrylate (PMMA) (495 A4 at 2000 RPM for 2 minutes) was spin-coated on the surface of the BST followed by hard-baking process at $180\degree$ for 90 seconds. Then, the PMMA-BST surface was metalized with 100 nm of Au deposited by dual electron-beam evaporation. Afterwards, a positive resit (AZ-1512 at 4000 RPM) of 1.5 $\mu$m was spin coated on the metalized surface followed by hard-baking process at $90\degree$ for 90 seconds.
By using a laser writer (MLA-100), the proposed SRRs (negative mask design) were patterned with light source of 365 nm,
expose dose energy of 150 mJ/cm$^2$ and DeFoc of 2 which is followed by hard-baking process at $110\degree$ for 60 seconds. The patterned samples were then developed in a developer (AZ-300 MIF for 2 minutes) followed by DI water rinse. Subsequently, the samples were etched using an ion mill (IntlVAC Nanoquest) at angle of $275\degree$ for 4 minutes with an etch rate of 30 nm/min. The active surface area of the fabricated device is $\approx$ 0.9 cm x 0.9 cm.\\ 

The PMMA layer serves two purposes.   First, it allowed us to tune the electromagnetic coupling strength of the SRR to the film.   Second, after TDTS measurements, it allowed the SRR layers to be removed cleanly in acetone via ultrasound.  As a result, it provided the possibility that the BST thin films can be re-used multiple times (up to 12 times) without any damage to their surface properties and leaving their phonon resonance intact.

    

\section{\bf{SD}: Time-Domain Terahertz Spectroscopy Measurements of BST-SRR Hybrid Systems}

To characterize the transmission spectra of the BST-SRRs, we employed the time-domain terahertz spectroscopy (TDTS) technique using a TOPTICA Teraflash system. The TOPTICA system uses fiber optic coupled InGaAs/InP-based photoconductive antennas (PCA) with bandwidth $>$ 6 THz instead of free space optics and LT-GaAs PCAs.
Both emitter and receivers are coupled to a laser with wavelength of 1560 nm, giving pulse width of 80 fs with 100 MHz repetition rate. The emitter is an InGaAs/InP PCA with 100 $\mu$m strip-line antenna and a receiver
is an InGaAs/InP PCA with 25 $\mu$m dipole antenna. It has two mechanical
delay stages: a long-travel delay stage comprising of corner-cube mirror which
introduces a constant timing offset (of up to 3000 ps), that compensates for any THz path length change e.g. shorter or longer optical fiber. Once the offset is fixed, the second highly precise but much faster delay stage is used for the actual time variation in the measurement. It consists of a voice-coil driven
corner-cube mirror combined with a digital, high-precision position sensor. The sensor records 50000 time stamps per second, with a resolution of 1.3 fs. These time stamps are synchronized with the readout of the signal values from the
terahertz receiver. Data acquisition is accomplished both during the forward
and backward movement of the mirror, which minimizes the “dead time”
of the system. Additionally, our TDTS system is equipped with a custom built cryogenic system that allows us to perform low temperature measurements up to 5 K.

\begin{figure}[h!]
    \includegraphics[width=0.4\columnwidth]{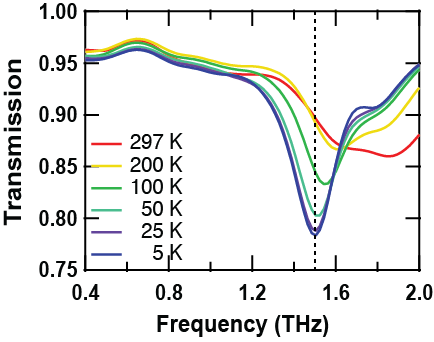}
    \caption{TDTS temperature dependent measurement of a bare (Bi$_x$Sb$_{1-x}$)$_2$Te$_3$ film showing the damping of the $\alpha$ phonon at higher temperature.}
    \label{SFig2}
    
\end{figure}

\begin{figure}[h!]
    \includegraphics[width=1\columnwidth]{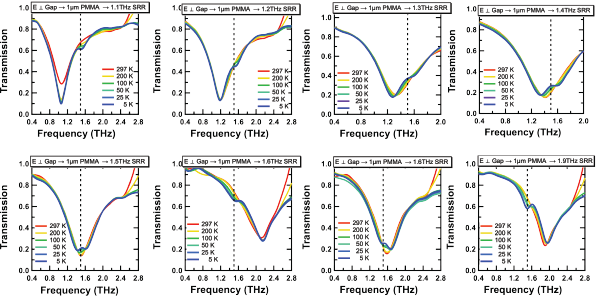}
    \caption{TDTS temperature dependent measurements of the BST-SRR hybrid systems with 1$\mu$m PMMA thickness at different SRR frequencies.}
    \label{SFig3}
    
\end{figure}

The temperature-dependent TDTS transmission measurement of a bare (Bi$_x$Sb$_{1-x}$)$_2$Te$_3$ film is illustrated in Fig. \ref{SFig2}. As temperature decreases, the phonon resonance becomes notably pronounced, with a resonance frequency  centered approximately around 1.5 THz. To induce strong coupling in a topological material via metamaterials, we fabricated seven different SRRs on 20 nm thick BST films, with a 1 $\mu$m PMMA layer serving as a protective coating. Fig. \ref{SFig3} displays the temperature-dependent transmission data of the coupled BST-SRR hybrid system for the various SRRs studied. At low temperature, it is apparent that the SRR resonance frequency dominates both below and above the phonon resonance frequency. However, as the SRRs approach the phonon resonance frequency, a splitting of the center frequency emerges. This split is attributed to the hybridization between the SRR resonant frequency and the $\alpha$ phonon mode of (Bi$_x$Sb$_{1-x}$)$_2$Te$_3$.  The temperature dependence here is interesting also.   In all cases, when the signature of the phonon disappears, the remaining single resonance appears at a frequency close to what the simulation predicts for the SRR. As samples are cooled down, a notable splitting emerges, but only for samples where the SRR frequency has been tuned close to the phonon.


\section{\bf{SE}: Double Lorentzian Model}

\begin{figure}
    \includegraphics[width=0.4\columnwidth]{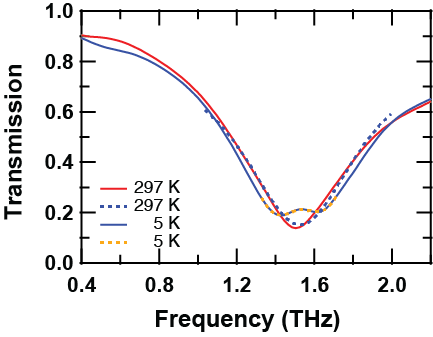}
    \caption{Sample double Lorentzian fit of TDTS temperature dependent measurement of BST-SRR hybrid system with resonance frequency $\approx$ 1.5 THz at room (\textit{red}) and low (\textit{blue}) temperatures.}
    \label{SFig4}
    
\end{figure}

We employed a double Lorentzian model to fit the transmission spectrum of the BST-SRR hybrid systems, as depicted in the Fig. \ref{SFig4}. The equation representing the double Lorentzian is presented in Eq. \ref{Seq1}, where A$_{0}$ denotes the amplitude and $\Gamma$ represents the half-width at half-maximum. From these fit parameters, we derived the resonance frequencies $\omega_-$, $\omega_+$, and damping rates $\Gamma_+$, $\Gamma_-$ for all SRR configurations. Both room temperature and low temperature data were subjected to fitting. 
The resonance frequencies of the SRRs measured at room temperature closely align with the simulated values. However, at lower temperatures, owing to the hybridization between the SRR resonant frequency and the $\alpha$ phonon mode of (Bi$_x$Sb$_{1-x}$)$_2$Te$_3$, we observe two eigenfrequencies corresponding to the lower and upper polaritonic branches. Detailed fit parameters extracted from these analyses are presented in Table \ref{Table 2}.

\begin{equation}
    \centering 
    L(\omega) = A_0 + \bigg [ A_+ \frac{ \frac {\Gamma_+}{2}}{(\omega - \omega_{+})^2 +\frac{\Gamma_+}{2}} + A_- \frac{  \frac {\Gamma_-}{2}}{(\omega - \omega_{-})^2 +\frac{\Gamma_-}{2}}\bigg]
     \label{Seq1}
 \end{equation}

\begin {table}
    \centering
\begin{tabular}{ |p{2cm}|p{2cm}|p{1.5cm}|p{1.5cm}|p{1.5cm}|p{1.5cm}|}
\hline
\hline
HFSS-SRR (THz) & SRR-297K (THz) & $\omega_-$5K (THz)  & $\Gamma_-$5K (THz) & $\omega_+$5K (THz) & $\Gamma_+$5K (THz)  \\
\hline
1.1 & 1.0601 &1.0536 &0.3322 &1.5327  &0.1387 \\
1.2 & 1.2237 &1.1742 &0.5001        &1.5669  &0.4368\\
1.3 & 1.3525 &1.2575 &0.4958        &1.6119  &0.4431\\
1.4 & 1.4201 &1.2952 &0.5468        &1.6244
&0.4891\\
1.5 & 1.5217 &1.3665 &0.5402        &1.6609  &0.5449\\
1.6 & 1.5991 &1.3917 &0.4639        &1.6946  &0.5985\\
1.7 & 1.7061 &1.4186 &0.3163        &1.7865  &0.6201\\
\hline
\end{tabular}

\caption{HFSS simulated and TDTS obtained resonant frequencies of the investigated split ring resonators along with the two hybrid polartionic modes obtained from the double Lorentzian fit of the TDTS low-temperature transmission spectrum.}
\label{Table 2}
\end{table}

\bibliography{suppl}